\documentclass{elsart}
\usepackage{psfig}
\def\la{\mathrel{\hbox{\rlap{\hbox{\lower4pt\hbox{$\sim$}}}\hbox{$<$}}}}
\def\ga{\mathrel{\hbox{\rlap{\hbox{\lower4pt\hbox{$\sim$}}}\hbox{$>$}}}}

\newcommand{\be}{\begin{eqnarray}}
\newcommand{\ee}{\end{eqnarray}}

\newcommand{\msol}{\ifmmode{{\rm M}_\odot}\else{M$_\odot$}\fi}
\newcommand{\foe}{\ifmmode{10^{51}}\else{$10^{51}$}\fi}

\newcommand{\xni}{\ifmmode{{\rm X}_{\rm Ni}}\else{X$_{\rm Ni}$}\fi}

\def\Teff{\ifmmode{T_{\rm eff}}\else{\hbox{$T_{\rm eff}$} }\fi}
\def\Tmod{\ifmmode{T_{\rm model}}\else{\hbox{$T_{\rm model}$} }\fi}
\def\Rzero{\ifmmode{R_0}\else{\hbox{$R_0$} }\fi}
\newcommand{\vno}{v_0}

\def\SP2{{\tt IBM SP2}}

\def\MPI{{\tt MPI}}

\def\PC2{{\tt PC$^2$}}

\def\inu{\ifmmode{I_{\nu}}\else{\hbox{$I_{\nu}$} }\fi}
\def\snu{\ifmmode{S_{\nu}}\else{\hbox{$S_{\nu}$} }\fi}
\def\jnu{\ifmmode{J_{\nu}}\else{\hbox{$J_{\nu}$} }\fi}

\def\fep{\ifmmode{{\rm Fe II}}\else\hbox{Fe~II }\fi}

  at 12.0 true pt 

\def\phoenix{{\tt PHOENIX}}

\def\phoenix{{\tt PHOENIX}}

\def\b{\beta}

\def\e{\epsilon}

\def\l{\lambda}
\def\L{\Lambda}
\def\t{\tau}

\def\div#1#2{{#1\over #2}}
\def\rout{\ifmmode{r_{\rm out}}\else\hbox{$r_{\rm out}$}\fi}
\def\tmax{\ifmmode{\tau_{\rm max}}\else\hbox{$\tau_{\rm max}$}\fi}
\def\tstd{\ifmmode{\tau_{\rm std}}\else\hbox{$\tau_{\rm std}$}\fi}
\def\vmax{\ifmmode{v_{\rm max}}\else\hbox{$v_{\rm max}$}\fi}
\def\muE{\ifmmode{\mu_{\rm E}}\else\hbox{$\mu_{\rm E}$}\fi} 
\def\pE{\ifmmode{p_{\rm E}}\else\hbox{$p_{\rm E}$}\fi} 
\def\bmax{\ifmmode{\b_{\rm max}}\else\hbox{$\b_{\rm max}$}\fi}
\def\kms{\hbox{$\,$km$\,$s$^{-1}$}}

\def\Teff{\hbox{$\,T_{\rm eff}$} }

\def\rout{\hbox{$r_{\rm out}$} }

\def\Pgas{\hbox{$P_{\rm gas}$} }
\def\chistd{\ifmmode{\chi_{\rm std}}\else\hbox{$\chi_{\rm std}$}\fi}

\def\msol{$M_\odot$}
\def\foe{10^{51}}

\def\xni{{\rm X}_{\rm Ni}}
\def\vno{v_0}

\def\lstar{\ifmmode{\Lambda^*}\else\hbox{$\Lambda^*$}\fi} 
\def\Rop{\ifmmode{[R_{ij}]}\else\hbox{$[R_{ij}]$}\fi}
\def\Rij{\Rop}
\def\Rji{\ifmmode{[R_{ji}]}\else\hbox{$[R_{ji}]$}\fi}
\def\Rstar{\ifmmode{[R_{ij}^*]}\else\hbox{$[R_{ij}^*]$}\fi}
\def\Rijstar{\Rstar}
\def\Rjistar{\ifmmode{[R_{ji}^*]}\else\hbox{$[R_{ji}^*]$}\fi}
\def\DRji{\ifmmode{[\Delta R_{ji}]}\else\hbox{$[\Delta R_{ji}]$}\fi}
\def\DRij{\ifmmode{[\Delta R_{ij}]}\else\hbox{$[\Delta R_{ij}]$}\fi}

\def\Jb{{\bar J}}
\def\Jnew{{\bar J_{\rm new}}}
\def\Jold{{\bar J_{\rm old}}}
\def\Jfs{{\bar J_{\rm fs}}}
\def\Snew{{S_{\rm new}}}
\def\Sold{{S_{\rm old}}}

\def\ns{\ifmmode{N_{\rm s}}          
        \else\hbox{$N_{\rm s}$}\fi}

\def\mat#1{{\bf #1}}     
\def\vek#1{{#1}}         

\newcount\eqcount
\def
  \nummer{
    \global\advance\eqcount by 1
    (\the\eqcount)
  }

\def
  \numadv{
    \global\advance\eqcount by 1
  }

\def
   \numout#1{
     (\the\eqcount #1)
  }

\def\ivek#1#2{\ifmmode{\vek{I}^{#1}_{#2}}
        \else\hbox{$\vek{I}^{#1}_{#2}$}\fi}

\def\tmat#1#2{\ifmmode{\mat{t}^{#1}_{#2}}
        \else\hbox{$\mat{t}^{#1}_{#2}$}\fi}
\def\rmat#1#2{\ifmmode{\mat{r}^{#1}_{#2}}
        \else\hbox{$\mat{r}^{#1}_{#2}$}\fi}
\def\bvek#1#2{\ifmmode{\beta^{#1}_{#2}}
        \else\hbox{$\beta^{#1}_{#2}$}\fi}

\def\lp{\ifmmode{\lambda^+_\tau}           
        \else\hbox{$\lambda^+_\tau$}\fi}
\def\lm{\ifmmode\lambda^-_\tau             
        \else\hbox{$\lambda^-_\tau$}\fi}

\begin{document}

\bibliographystyle{elsart}

\begin{frontmatter}
%
\title{Numerical Solution of the Expanding Stellar Atmosphere Problem}
%
\author{Peter H.~Hauschildt}
\address{Department of Physics and Astronomy \& Center for Simulational
Physics, University of Georgia, Athens, GA 30602-2451 USA
}
%
\author{E.~Baron}
\address{Department of Physics and Astronomy, University of
Oklahoma, 440 W. Brooks, Rm 131, Norman, OK 73019-0225 USA
}
%
%
\begin{center}
\em Full version available at \\
\small
\tt ftp://calvin.physast.uga.edu:/pub/preprints/CompAstro.ps.gz
\end{center}
%
\begin{abstract}
In this paper we discuss numerical methods and algorithms for the solution of
NLTE stellar atmosphere problems involving expanding atmospheres, e.g., found
in novae, supernovae and stellar winds. We show how a scheme of nested
iterations can be used to reduce the high dimension of the problem to a number
of problems with smaller dimensions. As examples of these sub-problems, we
discuss the numerical solution of the radiative transfer equation for
relativistically expanding media with spherical symmetry, the solution of the
multi-level non-LTE statistical equilibrium problem for extremely large model
atoms, and our temperature correction procedure. Although modern iteration
schemes are very efficient, parallel algorithms are essential in making large
scale calculations feasible, therefore we discuss some parallelization schemes
that we have developed.
\end{abstract}
\end{frontmatter}


\section{Introduction}

Astronomy is sometimes described as a ``passive science'' since it depends on
observations of distant objects, rather than on laboratory experiments. Due to
both laboratory measurements of important astrophysical atomic and nuclear
data, and advances in computational power which allow us to perform ``numerical
experiments'' that situation has changed in the last 50 years, and astronomy
has matured into the modern subject of astrophysics. Still, our ability to
understand the nature of astronomical objects is hampered by the fact that
astronomical observations detect radiation that has been emitted primarily from
the surface of objects. Thus in order to determine the structure of stellar
objects one must solve the radiation transport equation and compare ``synthetic
spectra'' with observations. The numerical solution of the radiation transport
problems is an important prerequisite for the calculation of model stellar
atmospheres. The simulated spectrum is then compared to observed spectra of objects
such as stars, where the radiation is mostly emitted from the outer layers. In
the case of very low mass stars and brown dwarfs, the atmosphere is also 
crucial in determining the interior structure of these objects since it
serves as a boundary condition for the equations of stellar structure in a nearly
fully convective ``star''.

Additionally, in objects such as supernovae, where the the explosion causes the
``atmosphere'' (here used to paraphrase ``the region where the spectrum
forms'') to expand rapidly, a time series of spectra reveal the entire
structure of the object as the ejected material expands and thins and the
atmosphere moves inward in the material.  In the case of expanding objects such
as hot stars (many with strong stellar winds), novae, and supernovae; the
radiation transport equation must be solved simultaneously with the
hydrodynamical equations, an even more difficult computational problem than
static stars.  We focus here on the computation of model atmospheres and the
numerical solution of the radiation transport equation in expanding media with
known velocity fields. This is a frequently encountered situation, e.g., when
the hydrodynamic behavior is known a priori, or can be calculated separately
from the radiation transport by using a nested iteration scheme.  The feedback
between detailed synthetic spectrum calculations and hydrodynamic simulations
is often the primary tool for testing a specific hydrodynamical model.

Our group has developed the very general non-LTE (NLTE) stellar
atmosphere computer 
code {\tt PHOENIX}
\cite{phhs392,phhcas93,hbfe295,phhnov95,faphh95,phhnovfe296,snefe296,hbapara97,bhpar298}
which can handle very large model atoms as well as line blanketing by hundreds
of millions of atomic and molecular lines.  This code is designed to be both
portable and very flexible: it is used to compute model atmospheres and
synthetic spectra for, e.g., novae, supernovae, M and brown dwarfs, O to M
giants, white dwarfs and accretion disks in Active Galactic Nuclei (AGN). The
radiative transfer in \phoenix\ is solved in spherical geometry and includes
the effects of special relativity (including advection and aberration) in the
modeling. 
The \phoenix\ code allows us to include a large number of
NLTE and LTE background spectral lines and solves the radiative transfer
equation for each of them {\em without} using simple approximations
like the Sobolev approximation. Therefore, the profiles of spectral lines must
be resolved in the co-moving (Lagrangian) frame. This requires many
wavelength points (we typically use 150,000 to 300,000 points).  Since the CPU
time scales linearly with the number of wavelength points, the CPU time
requirements of such calculations are large. In addition, (NLTE)
radiative rates for both line and continuum transitions must be calculated
and stored at every spatial grid point for each transition, which requires
large amounts of storage and can cause significant performance degradation
if the corresponding routines are not optimally coded.

In strict LTE the radiation and matter are assumed to be in
equilibrium with each 
other everywhere throughout the atmosphere. In LTE the source function
is assumed to be given by the Planck function (see below). In NLTE,
the radiation is no longer assumed to be in equilibrium with the
matter and hence the full coupling between matter and radiation must
be calculated in order to calculate the source function.

We concentrate here on the calculation of model atmospheres for
expanding media and, in addition, describe some of the important parts of the
numerous numerical algorithms used in \phoenix: the numerical solution of the
radiation transport equation, the non-LTE rate equations, and the
parallelization of the code. An important problem in these calculations
is to find a consistent solution of the very diverse equations that
describe the various physical processes. We have developed a scheme of
nested iterations that enables us to separate many of the variables
(e.g., separating the temperature correction procedure from the 
calculation of the NLTE occupation numbers). This allows us to 
compute far more detailed stellar atmosphere models than was previously
possible. We will give an outline of these methods in this paper.

In order to take advantage of the enormous computing power and vast aggregate
memory sizes of modern parallel supercomputers, both potentially allowing much
faster model construction as well as more sophisticated models, we have
developed a parallel version of {\tt PHOENIX}. Since the code uses a modular
design, we have implemented different parallelization strategies for different
modules (e.g., radiative transfer, NLTE rates, atomic and molecular line
opacities) in order to maximize the total parallel speed-up of the code. In
addition, our implementation allows us to change the distribution of
computational work onto different nodes both via input files and dynamically
during a model run, which gives a high degree of flexibility to optimize
performance for both a number of different parallel supercomputers (we are
currently using {\tt IBM SP2}s, {\tt SGI Origin 2000}s, {\tt HP/Convex
SPP-2000}s, and {\tt Cray T3E}s) and for different model parameters.  Since
\phoenix\ has both large CPU and memory requirements we have developed the
parallel version of the code using a MIMD approach.  We use the \MPI\ message
passing library \cite{mpistd} for portability and {\em simultaneously} use both
task and data parallelism in order to optimize the parallel speed-up
\cite{hbapara97,bhpar298}.

\section{The Problem}

\subsection{Overview}

Our goal (for the purposes of this paper) is to construct self-consistent
models of expanding stellar atmospheres.  The atmosphere itself is
parameterized by a number of parameters, e.g., the total energy emitted by the
object (luminosity $L$), the mass $M$ of the star, the abundances of the
elements (in some cases as function of the location in the atmosphere).  This
means that we have to find a set of physical variables such as temperatures,
densities, population number of each atomic energy level and the radiation
field, at {\em each} location in the atmosphere so that all constraint
equations are simultaneously fulfilled.  In Fig.~\ref{flow1} we show this
requirement in a simplified graphical form where the arrows indicate {\em
direct} (usually non-linear) coupling between the different blocks. The number
of variables that need to be addressed is, formally, very large. A typical case
of a spherically symmetric shell model with 50 radial points requires a set of
50 temperatures and gas pressures (or matter densities). In addition to this we
include a set of about 6000 individual energy levels for atoms and ions
directly, the population of each must be known at every radial point, adding a
total of 300,000 variables. In order to calculate the population numbers, we
need a description of the radiation field at each radial point and on a set of
wavelength points (the rates that govern the transitions between atomic energy
levels are integrals of the mean intensity of the radiation field over
wavelength). We typically need 100,000 to 300,000 wavelength point to describe
the complete radiation field, which adds, in the worst case, 15 million
variables to the system. 

Fortunately, most of these formal variables are tightly coupled to a much
smaller set of variables which we might, therefore, consider the
``fundamental'' variables of the model atmosphere problem. In our approach,
these fundamental variables are the temperatures $T$, the gas pressures
$\Pgas$, and the population numbers $n_i$ at each radial point $r_i$. The
radiation field is considered a ``derived'' quantity and the problem is thus
reduced to find a set of physical variables $\{T,\Pgas, n_i\}$ at each radial
point $i$ so that the system outlined in Fig.~\ref{flow2} is self-consistent.
With this approach we have reduced the number of variables from several million
to a few 100 thousand, which is still a daunting number.

Although it is possible to analytically bring the system into a form so that it
could be solved by a Newton-Raphson approach \cite{mihalas78sa}, this idea is
computationally prohibitive because of its enormous memory and time
requirements (however, for smaller systems this approach has been used
successfully). Furthermore, this approach is complex to implement and it is
relatively hard to add more ``physics'' to the model atmosphere. We have thus
developed a scheme of nested iterative solutions that considers the direct (or
strong) couplings between important variables directly and iteratively accounts
for the indirect coupling between sets of variables. With this approach the
problem of constructing the model atmosphere can be separated into solving a
large number of smaller problems with only a few 100 variables.  The global
requirement of a self-consistent solution is then reach by iteratively coupling
these sets of variables to each other until a prescribed accuracy has been
reached. This method works because the level of coupling between the variables
is very different. For example, the temperature structure of the atmosphere
depends mostly on the global constraint of energy conservation (represented by
wavelength integrals over the whole spectrum) and on the {\em ratios} of
several averaged opacities, but it does not depend strongly on the fine {\em
details} of the radiation field or the {\em individual} population of the vast
majority of the atomic levels. Therefore, {\em correction} to the temperature
structure can be calculated approximately. The current errors of, e.g., the
energy conservation equations, must be calculated exactly in order to this
scheme to function, however, this is relatively simple. The general idea of
calculating errors exactly but corrections to the variables approximately will
work if the approximations are good enough for the scheme to converge at all.
This method will require more iterations to reach convergence but this is more
than offset by faster individual iterations and (very often) by better
robustness. The latter is very important if many model atmospheres have to be
constructed or if no good initial guesses for the the variables are known.

In the following sections we will concentrate on a few key parts of the
expanding atmosphere problem: the numerical solution of the radiative transfer
equation for a single wavelength point (this will deliver the radiation field
for a given set of variables at every wavelength), the solution of the NLTE
statistical equilibrium equations (coupling the radiation field to the level
populations), and an outline of the temperature correction procedure. The
latter is important because it allows us to solve the NLTE statistical
equilibrium equations separately for individual elements (and even ionization
stages), which dramatically reduces the dimension of the sub-problems that have
to be solved within the global nested iteration scheme. In this paper we will
not discuss problems related to the hydrodynamics of the expanding medium 
or the details of the equation of state calculations, both of which are 
important topics.

\subsection{Radiative transfer in expanding media}

The equation of radiative transfer (RTE) in spherical symmetry for moving media
has been solved with a number of different methods, e.g.\ Monte Carlo
calculations \cite{magnan70,cns72,avb72}, Sobolev methods \cite{castor70}, the
tangent ray method \cite{mkh75}, and the DOME method \cite{phhrw91}. Only the
tangent ray and the DOME method have been used to solve the RTE for very fast
expanding shells (e.g.\ supernovae or novae) including the necessary special
relativistic terms.  Both methods need relatively large amounts of CPU time to
compute the radiation field, mainly because of the need for matrix inversions
(tangent ray method) or matrix diagonalization (DOME), which make both of them
impractical for use within radiation-hydrodynamic studies of nova or supernova
explosions.  It has been shown \cite{hbw91} that the special relativistic terms
in the RTE can be very important, even in the optically thick regions of
expanding shells, and lead to results different than from the simpler approach
which simply neglects the relativistic terms.

Recently, iterative methods for the solution of the RTE have been developed,
based on the philosophy of operator perturbation \cite{cannon73,scharmer84}.
Following these ideas, different approximate $\Lambda$-operators for this
``accelerated $\Lambda$-iteration'' (ALI) method have been used successfully
\cite{OAB,hamann87,werner87} and have been applied to the construction of
non-LTE, radiative equilibrium models of stellar atmospheres \cite{werner87}.

We describe the use of the short-characteristic method \cite{OAB,ok87} to
obtain the formal solution of the special relativistic, spherically symmetric
radiative transfer equation (SSRTE) along its characteristic rays and then use
a band-diagonal approximation to the discretized $\Lambda$-operator
\cite{phhs392,ok87,hsb94} as our choice of the approximate $\Lambda$-operator.
This method can be implemented very efficiently to obtain an accurate solution
of the SSRTE for continuum and line transfer problems using only modest amounts
of computer resources.

The co-moving frame radiative transfer equation for spherically
symmetric flows can be written as \cite{found84}:

\vbox{\be
&\quad&\gamma (1+\beta\mu)\frac{\partial\inu}{\partial t} + \gamma (\mu +
\beta) \frac{\partial\inu}{\partial r}\nonumber\\
& +& \frac{\partial}{\partial
\mu}\left\{ \gamma (1-\mu^2)\left[ \frac{1+\beta\mu}{r}
\right.\right.\nonumber\\
&\quad&\left.\left. \quad -\gamma^2(\mu+\beta)
\frac{\partial\beta}{\partial r} -  
\gamma^2(1+\beta\mu) \frac{\partial\beta}{\partial
t}\right] \inu\right\} \nonumber\\
&-&  \frac{\partial}{\partial
\nu}\left\{ \gamma\nu\left[ \frac{\beta(1-\mu^2)}{r}
+\gamma^2\mu(\mu+\beta) \frac{\partial\beta}{\partial r}
\right.\right.\nonumber\\
&\quad& \left.\left.\quad  +
\gamma^2\mu(1+\beta\mu) \frac{\partial\beta}{\partial
t}\right]\inu\right\}\label{fullrte}\\
&+&\gamma\left\{\frac{2\mu+\beta(3-\mu^2)}{r}\right.
\nonumber\\
&\quad&\quad
\left. +\gamma^2(1+\mu^2+2\beta\mu)\frac{\partial\beta}{\partial r} + 
\gamma^2[2\mu + \beta(1+\mu^2)]\frac{\partial\beta}{\partial
t}\right\}\inu \nonumber\\
&\quad& = \eta_\nu - \chi_\nu\inu.\nonumber
\ee}
\noindent
$\beta=v/c$ is the velocity in units of the speed of light, $c$; and
$\gamma = (1-\beta^2)^{-1/2}$ is the usual Lorentz factor. 
Equation~\ref{fullrte} is a integro-differential equation, since the
emissivity $\eta_\nu$ contains \jnu\unskip, the zeroth angular moment of
\inu\unskip:
\[ \eta_\nu = \kappa_\nu \snu + \sigma_\nu \jnu 
 +\sum_{\rm lines} \sigma_l(\nu) \int \phi_l J_\nu \,d\nu, \]
with 
\be
\jnu = 1/2 \int_{-1}^{1} d\mu\, \inu,\label{jdef}
\ee
where $\snu$ is the source function, $\kappa_\nu$ is the absorption
coefficient, $\sigma_\nu$ is the scattering coefficient for continuum 
processes,  $\sigma_l$ are the line scattering coefficients, and
$\phi_l$ is the line profile function. The independent variables
are the radius $r$ of the shell, the cosine $\mu$ of the angle 
between the radial direction and the propagation vector of the 
light (with $\mu=-1,1$ for radially inward and outward moving light, 
respectively), and the frequency $\nu=c/\lambda$ for a wavelength $\l$
of the light.
With
the assumption of time-independence, $\frac{\partial\inu}{\partial t} =
0$, and a monotonic velocity field Eq.~\ref{fullrte} becomes a boundary-value problem in
the spatial coordinate and an initial value problem in the frequency
or wavelength coordinate.

Switching from frequency to wavelength (Eq.~\ref{fullrte} is presented
in Ref.~\cite{phhs392} in wavelength),
the mean intensity $J_\lambda$ is obtained from the source function
$S_\lambda$ by a formal solution of the RTE which is symbolically written
using the $\Lambda$-operator $\Lambda_\lambda$ as
\begin{equation}
     J_\lambda = \Lambda_\lambda S_\lambda.              \label{frmsol}
\end{equation}
In the case of the transition of a two-level atom, we have
\be
     \Jb = \L S,           \label{etla}
\ee
where $\bar J=\int \phi(\lambda) J_\lambda \,d\lambda$, 
$\Lambda=\int \phi(\lambda) \Lambda_\lambda \,d\lambda$ with the normalized
line profile $\phi(\lambda)$. The line source function, for the simple case
of a two-level atom without continuum and background absorption or
scattering, is given by $S=(1-\e)\Jb + \e B$, where $\e$ 
denotes the thermal coupling parameter and $B$ is Planck's function.

The $\L$-iteration method, i.e.\ to solve Eq.~\ref{etla} by a fixed-point
iteration scheme of the form
\be
   \Jnew = \L \Sold , \quad
   \Snew = (1-\e)\Jnew + \e B  ,\label{alisol}
\ee
fails in the case of large optical depths and small $\e$. This result is caused
by the fact that the largest eigenvalue of the amplification matrix (in the
case of Doppler-profiles) is approximately \cite{mkh75}
$\l_{\rm max} \approx (1-\e)(1-T^{-1})$, where $T$ is the  optical 
thickness of the medium. For small $\e$ and large $T$, this is very close
to unity and, therefore, the convergence rate of the $\L$-iteration is very 
poor. A physical description of this effect can be found in 
Mihalas \cite{mih80}.

\subsubsection{The operator splitting method}

The idea of the ALI or operator splitting method is to reduce the 
eigenvalues of the amplification matrix in the iteration scheme
\cite{cannon73}  by 
introducing an approximate $\L$-operator (ALO) $\lstar$
and to split $\L$ according to
\be
           \L = \lstar +(\L-\lstar) \label{alodef}
\ee
and rewrite Eq.~\ref{etla} as
\be
     \Jnew = \lstar \Snew + (\L-\lstar)\Sold. 
\ee
This relation can be written as \cite{hamann87}
\be
    \left[1-\lstar(1-\e)\right]\Jnew = \Jfs - \lstar(1-\e)\Jold, \label{alo}
\ee
where $\Jfs=\L\Sold$. Equation~\ref{alo} is solved to get the new values of 
$\Jb$ which is then used to compute the new 
source function for the next iteration cycle.

Mathematically, the ALI method belongs to the same family of iterative
methods as the Jacobi or the Gauss-Seidel methods
\cite{golub89:_matrix}. These 
methods have the general form
\be
    M x^{k+1} = Nx^{k} + b
\ee
for the iterative solution of a linear system $Ax=b$ where the system
matrix $A$ is split according to $A=M-N$. In the case of the ALI
method we have $M=1-\lstar(1-\e)$ and, accordingly,
$N=(\L-\lstar)(1-\e)$ for the system matrix $A=1-\L(1-\e)$. The
convergence of the iterations depends on the spectral radius,
$\rho(G)$, of the iteration matrix $G=M^{-1}N$.  For convergence the
condition $\rho(G)<1$ must be fulfilled, this puts a restriction on
the choice of $\lstar$. In general, the iterations will converge
faster for a smaller spectral radius.  To achieve a significant
improvement compared to the $\L$-iteration, the operator $\lstar$ is
constructed so that the eigenvalues of the iteration matrix $G$ are
much smaller than unity, resulting in swift convergence. Using
parts of the exact $\L$ matrix (e.g., its diagonal or a tri-diagonal
form) will optimally reduce the eigenvalues of the 
$G$. The
calculation and the structure of $\lstar$ should be simple in order to
make the construction of the linear system in Eq.~\ref{alo} fast. For
example, the choice $\lstar=\L$ is best in view of the
convergence rate (it is equivalent to a direct solution by matrix inversion)
but the explicit construction of $\L$ is more time
consuming than the construction of a simpler $\lstar$. The solution of
the system Eq.~\ref{alo} in terms of linear algebra, using modern
linear algebra packages such as, e.g., {\tt LAPACK}, is so fast that
its CPU time can be neglected for the small number of variables
encountered in 1D problems (typically the number of discrete shells
is about 50). However, for 2D or 3D problems the size of $\L$ gets
very large due to the much larger number of grid points as compared to
the 1D case.  Matrix inversions, which are necessary to solve
Eq.~\ref{alo} directly, therefore become extremely time
consuming. This makes the direct solution of Eq.~\ref{alo} more CPU intensive
even for $\lstar$'s of moderate bandwidth, except for the trivial case
of a diagonal $\lstar$.  Different methods like modified conjugate
gradient methods \cite{turek93} may be effective for these
2D or 3D problems.

The CPU time required for the solution of the RTE using the ALI method depends
on several factors: (a) the time required for a formal solution and the
computation of $\Jfs$, (b) the time needed to construct $\lstar$, (c) the time
required for  the solution of Eq.~\ref{alo}, and (d) the number of iterations
required for convergence to the prescribed accuracy. Points (a), (b) and (c)
depend mostly on the number of discrete shells, and can be assumed to be fixed
for any given configuration. However, the number of iterations required to
convergence depends strongly on the bandwidth of $\lstar$.
This indicates, that there is an {\em optimum
bandwidth} of the $\lstar$-operator which will result in the shortest possible
CPU time needed for the solution of the RTE, which we will discuss below.

\subsubsection{Computation of $\lstar$}

The formal solution of the SSRTE is performed along the characteristic
rays on a mesh $\{r_i\}$, $i=1,\ldots,N_s$ of discrete shells using the
short-characteristic (SC) method of Olson and Kunasz \cite{ok87} with
piece-wise parabolic or linear interpolation.  The characteristic rays are
{\em curved} in the case of the SSRTE and have to be calculated before the
solution of the radiative transfer equation proceeds (see \cite{phhs392}
for details).  Improvements in this method \cite{phhs392,hsb94} include
an improved angle integration using generalized Simpson-quadrature and
a generalization of the approximate $\L$-operator to an arbitrary number
of bands below and above the main diagonal (up to the full $\L$-operator).

We describe here the general procedure of calculating the $\lstar$ with
{\em arbitrary} bandwidth, up to the full $\Lambda$-operator, for the
SC method in spherical symmetry \cite{hsb94}. Although we consider the
SSRTE as given in the previous section, the same procedure applies for
radiative transfer problems in static media or in (static or moving)
media with plane-parallel symmetry. The specialization of the formulae
given in this section is straightforward.

The formal solution along a characteristic of the SSRTE (hereafter, a
``ray'') is done using a polynomial interpolation of the source
function, $S$, along the ray. For reasons of numerical stability, we
use linear or quadratic interpolation of $S$ along each ray,
although this is not required by the method. This leads to the
following expressions for the specific intensity $I(\tau_i)$ along a
ray (cf.~Ref.~\cite{ok87} for a derivation of the formulae):
\be
       I^k(\tau^k_i) &=& I^k(\tau^k_{i-1}) \exp(\tau^k_{i-1}-\tau^k_i)
                          +\int_{\tau^k_{i-1}}^{\tau^k_i} \hat S(\tau)
                                               \exp(\tau^k_{i-1}-\tau)
                                                          \, d\tau,\\
       I^k_i &\equiv& I^k_{i-1}\exp(-\Delta\tau^k_{i-1})+\Delta I^k_i
,\nonumber 
\ee
where the superscript $k$ labels the ray; $\tau^k_i$ denotes the
optical depth along the ray $k$ with $\tau^k_1\equiv 0$ and 
$\tau^k_{i-1} \le \tau^k_i$
while $\tau^k$ is calculated, e.g., using piecewise linear interpolation of
$\hat\chi$ along the ray, viz.\
\be
     \Delta\tau^k_{i-1} = (\hat\chi_{i-1}+\hat\chi_i)|s^k_{i-1}-s^k_i|/2.
\ee
and
\be
    \Delta I^k_i = \alpha^k_i \hat S_{i-1}
                    + \beta^k_i \hat S_i + \gamma^k_i \hat S_{i+1},
\ee
where $i$ is the ``running'' index along the ray and $|s^k_{i-1}-s^k_i|$ is 
the geometrical path length between points $i$ and $i-1$.
The expressions for the coefficients $\alpha^k_i$, $\beta^k_i$ and $\gamma^k_i$
are given in Ref.~\cite{ok87} (see also Ref.~\cite{phhs392}). 

We describe the  construction of $\lstar$ for arbitrary bandwidth using
the example of a characteristic that is tangential to an arbitrary shell:
Ray $k$ is the ray that is
tangent to shell $k+1$
The intersection points (including the point of
tangency) are labeled from left to right, the direction in which the
formal solution proceeds.   Ray $k$ has $2k+1$ points of
intersection with discrete shells $1\ldots k+1$.  To compute row $j$ of
the discrete $\Lambda$-operator (or $\Lambda$-matrix), $\Lambda_{ij}$,
we sequentially label the intersection points of the ray $k$ with the
shell $i$, and define auxiliary
quantities $\lambda_{ij}^k$ and $\hat\lambda_{ij}^k$ as follows:

\vbox{\be
     \lambda^k_{i,j} &= 0  &\quad {\rm for\ }  i<j-1  \nonumber\\
     \lambda^k_{j-1,j} &= \gamma^k_{j-1}   &\quad {\rm for\ }  i=j-1 \nonumber \\
     \lambda^k_{j,j} &= \lambda^k_{j-1,j}\exp(-\Delta\tau^k_{j-1}) +
\beta^k_{j}  
                                        &\quad{\rm for\ }  i=j  \nonumber\\
     \lambda^k_{j+1,j} &= \lambda^k_{j,j}\exp(-\Delta\tau^k_{j}) + 
\alpha^k_{j+1} 
                                        &\quad{\rm for\ }  i=j+1  \\
     \lambda^k_{i,j} &= \lambda^k_{i-1,j}\exp(-\Delta\tau^k_{i-1}) 
                                        &\quad{\rm for\ }  j+1 < i \le k+1 \nonumber
\ee}
For the calculation of $\hat\lambda^k_{i,j}$, we obtain:
\be
     \hat\lambda^k_{i,j} &= \lambda^k_{i-1,j}\exp(-\Delta\tau^k_{i-1}) 
                                       &\quad{\rm for\ }  i = k+2  \nonumber \\
     \hat\lambda^k_{i,j} &= \hat\lambda^k_{i-1,j}\exp(-\Delta\tau^k_{i-1}) 
                                        &\quad{\rm for\ }  k+2 <i < k+j+2 \nonumber  \\
     \hat\lambda^k_{i,j} &= \hat\lambda^k_{i-1,j}\exp(-\Delta\tau^k_{i-1})
                          + \alpha^k_i 
                                        &\quad{\rm for\ }  i = k+j+2   \nonumber\\
     \hat\lambda^k_{i,j} &= \hat\lambda^k_{i-1,j}\exp(-\Delta\tau^k_{i-1})
                          + \beta_i 
                                        &\quad {\rm for\ }  i = k+j+3   \nonumber\\
     \hat\lambda^k_{i,j} &= \hat\lambda^k_{i-1,j}\exp(-\Delta\tau^k_{i-1})
                          + \gamma^k_i 
                                        &\quad {\rm for\ }  i = k+j+4   \\
     \hat\lambda^k_{i,j} &= \hat\lambda^k_{i-1,j}\exp(-\Delta\tau^k_{i-1}) 
                          &\quad {\rm for\ }  k+j+5 \le i \le 2k+1   \nonumber
\ee

\noindent Using the $\lambda^k_{ij}$ and $\lambda^k_{ij}$, we can now write the
$\L$-Matrix as
\be
    \Lambda_{ij} = \sum_k \left( \sum_{\{l\}} w^k_{l,j}\lambda^k_{l,j}
                               + \sum_{\{l'\}}
w^k_{2(k+1)-l',j}\hat\lambda^k_{2(k+1)-l',j} 
                           \right),
\ee
where $w^k_{i,j}$ are the angular quadrature weights, $\{l\}$ is the set $\{i
\le k+1\}$ and $\{l'\}$ is the set $\{i > k+1\}$.
This expression gives the {\sl full} $\L$-matrix, it can easily be 
specialized to compute only {\sl certain bands} of the $\L$-matrix. In that  case, not all 
of the $\lambda^k_{i,j}$ and  $\hat\lambda^k_{i,j}$ have to be computed, 
reducing the CPU time from that required for
the computation of the full $\L$-matrix. 

\subsubsection{Numerical Considerations}

The calculation of $\lstar$ using the 
algorithm outlined can be vectorized and parallelized with respect to the 
ray index $k$ and the row index $j$ for any given bandwidth of $\lstar$. 
In addition, quantities like $\exp(-\Delta\tau^k_{i-1})$, $\alpha^k_i$, 
$\beta^k_i$ and $\gamma^k_i$ can be pre-calculated 
and stored, a process which is fully vectorizable and parallelizable. 

For each point on a ray, the computation of the specific intensity uses
about 7 floating point operations (flops), whereas the computation
of the $\lambda^k_{i,j}$ and  $\hat\lambda^k_{i,j}$ takes only 1
flop {\em per intersection point}. In addition, about 3 flops are
needed for the integration over the angle coordinate $\mu$ in order to
compute the mean intensities $J$ and the $\lstar$-operator.  We have
to calculate the formal solution for $ N_{\rm T}(N_{\rm T}+1)+N_{\rm
T}+2N_{\rm S}N_{\rm C}$ points, where $N_{\rm S}$ is the number
of discrete shells, $N_{\rm C}$ is the number of core intersecting
characteristics and $N_{\rm T}=N_{\rm S}-1$ is the number of tangent
rays. Therefore, the number of flops required for the computation of the
specific intensities at all points is $\approx 10[(N_{\rm S}+1)(N_{\rm
S}-1)+2N_{\rm S}N_{\rm C}]$.  To estimate the number of flops required
for the calculation of a $\lstar$-operator with a bandwidth of $N_{\rm
B} \le N_{\rm S}$, we assume that each point of a ray has $N_{\rm B}$
nearest neighbors, thus {\em overestimating} the number of operations. In
this approximation, we have to compute $\le N_{\rm B}N_{\rm T}(N_{\rm
T}+2)+2N_{\rm B}N_{\rm S}N_{\rm C}$ auxiliary variables $\lambda^k_{i,j}$
or $\hat\lambda^k_{i,j}$. Therefore, about $\le 4N_{\rm B}[(N_{\rm
S}-1)(N_{\rm S}+1)+2N_{\rm S}N_{\rm C}]$ floating point operations are
needed to compute the $\lstar$-operator and the ratio of the numerical
work needed for the computation of a $\lstar$-operator with a bandwidth
of $N_{\rm B}$ and one formal solution is of the order of $2N_{\rm
B}/5$. This expression actually {\sl significantly} overestimates the
number of operations required for the construction of the $\lstar$
operator, in particular for larger bandwidths (the effects of the
boundaries become more important for larger bandwidths). For example,
according to this estimate the computation of the full $\L$-matrix
for $N_{\rm S}=50$ takes about the same time as $20$ formal solutions,
however, the actual time used for the construction of the full $\L$-matrix
corresponds only to about $6$ formal solutions on many machines. This
indicates that the number of iterations must be rather small in order
to make ALO's with small bandwidth competitive in terms of speed for the
solution of radiative transfer problems and that the initial guess for the
source function will have a large influence on the optimum bandwidth. The
best strategy is to use monitoring to predict the ``optimum'' bandwidth
that gives the shortest time for the solution of the SSRTE at any given
wavelength point in an ``adaptive bandwidth operator splitting'' method,
see Ref.~\cite{hsb94} for details and results for a number of machines.

In order to accelerate convergence the Ng method \cite{ng74} or the
Orthomin method \cite{vinsome76} may be used (see Auer \cite{auer91}
for a review of different acceleration methods). These methods can cut
down the number of iterations required to reach a prescribed accuracy
by a factor of two or more with only a small increase in computational
overhead.

\subsection{NLTE calculations}

In order to solve Eq.~\ref{fullrte}, the emissivity $\eta_{\lambda}$
must be known, but $\eta_{\lambda}$ depends on the NLTE level
populations and therefore the NLTE rate equations must be solved
simultaneously with Eq.~\ref{fullrte}. This is further complicated by
the fact that the NLTE rate equations depend on the radiation field itself.
The NLTE rate equations have the form \cite{mihalas78sa}
\be
    \sum_{j<i} n_j \left(R_{ji}+C_{ji}\right)
   &-n_i\left\{\sum_{j<i} \left(\div{n_j^{*}}{n_i^{*}}\right)
		\left(R_{ij}+C_{ji}\right)
	     +\sum_{j>i} 
		\left(R_{ij}+C_{ij}\right)\right\} \label{rateeqs}  \\
    &\qquad +\sum_{j>i} n_j 
     \left(\div{n_i^{*}}{n_j^{*}}\right)\left(R_{ji}+C_{ij}\right)
     = 0. \nonumber
\ee
In Eq.~\ref{rateeqs}, $n_i$ is 
the actual, non-LTE population density of a level $i$
and the symbol $n_i^{*}$ denotes the so-called LTE population density
of the level $i$, which is given by
\be
     {n_i^{*}}
      = \div{g_i}{g_\kappa} {n_\kappa}
       {2 h^3 n_e  \over (2 \pi m)^{3/2} (kT)^{3/2}} \label{saha}
       \exp\left(-\div{E_i-E_\kappa}{kT}\right).
\ee
Here $n_\kappa$ denotes the {\it actual}, i.e., non-LTE,
population density of the ground
state of the next higher ionization stage of the same element;
${g_i}$ and ${g_\kappa}$ are
the statistical weights of the levels $i$ and $\kappa$, respectively.
In Eq.~\ref{saha}, $E_i$ is the excitation energy of the level $i$ and
$E_\kappa$ denotes the ionization energy from the ground state to 
the corresponding ground state of the next higher ionization stage.
The actual, non-LTE electron density is given by $n_e$. The system of
rate equations is closed by the conservation equations for the nuclei and the 
charge conservation equation (cf.\ Ref.~\cite{mihalas78sa}).

The sums in Eq.~\ref{rateeqs} extend only over the levels that are
included in our model atoms; for example, in singly ionized iron our
model atom
consists of 675 energy levels \cite{hbfe295}. The weaker radiative
transitions are treated as LTE background opacity (see
Refs.~\cite{phhcas93,hbfe295}). 

The rate coefficients for radiative and collisional transitions between
two levels 
$i$ and $j$ (including transitions from and to the continuum, see below)
are given by
$R_{ij}$ and $C_{ij}$, respectively. 
In our notation, the upward (absorption) radiative rate coefficients $R_{ij}$ 
($i<j$) 
are given by
\be
   R_{ij} = 
    {4\pi\over hc}
     \int_0^\infty \alpha_{ij}(\lambda)
J_\lambda(\lambda)\,\lambda d\lambda, 
\ee
whereas the downward (emission) radiative rate coefficients $R_{ji}$
($i<j$) are  
given by
\be
   R_{ji}= 
    {4\pi\over hc}
     \int_0^\infty \alpha_{ji}(\lambda)
     \left( {2hc^2\over\lambda^5} +J_\lambda(\lambda)\right) 
     \exp\left(-{hc\over k\lambda T}\right)\,\lambda d\lambda.
\ee
Here, $J$ is the mean intensity, $T$ the electron temperature, $h$ and
$c$ and Planck's constant and the speed of light, respectively. 
For the purposes of this paper, we assume that 
cross section $\alpha_{ij}(\lambda)$ of the transition $i\to j$ at the
wavelength $\lambda$ is known for both line and continuum transitions
and that it is the same for both absorption and emission processes
(complete redistribution).

Not all atomic processes fit neatly into the above scheme where the
rates are in detailed balance. Non-thermal ionization by fast
electrons, K-capture, Auger emission, and two-photon decay are
important in various stages of the evolution of novae and
supernovae. They can be included in the above formulation with
reasonable approximations, however.

\subsubsection{The Rate Operator}

As described above, a simple fixed point iteration scheme for the solution of
the rate equations will converge much too slowly to be useful for most cases of
practical interest. Therefore, we use an extension of the operator 
splitting idea for the solution of the rate equations.

We rewrite the rate equations in the form of an ``operator
equation.'' This equation is then used to introduce an ``approximate
rate operator'' in analogy to the approximate $\L$-operator which can
then be  used to iteratively solve the rate and statistical equations
by an operator splitting method, details of the approach 
are given in \cite{phhcas93}.

We introduce first the ``rate operator'' $[R_{ij}]$ for upward transitions
in analogy to the $\Lambda$-operator. $[R_{ij}]$ is defined so that
\be
 R_{ij} = [R_{ij}][n]. 
\ee
Here, $[n]$ denotes the ``population density operator'', which 
can be considered as the vector of the population densities of all 
levels at all points in the medium under consideration. 
The radiative rates are (linear) functions of the mean intensity $J$, 
which is given by
$J(\lambda) = \Lambda(\lambda) S(\lambda)$, where 
$S=\eta(\lambda)/\chi(\lambda)$ 
is the source function.
Using the $\Lambda$-operator, we can write $[R_{ij}][n]$ as:
\be
    [R_{ij}][n]={4\pi\over hc}
       \int \alpha_{ij}(\lambda)\Lambda(\l) S(\lambda) \,\lambda d\lambda.
\ee
This can be brought into the form (see \cite{phhcas93} for details)
\be
     \Rop [n] = {4\pi\over hc}
     \left[ \int_0^\infty \alpha_{ij}(\lambda)
	 \Psi(\lambda) E(\lambda) \,\lambda d\lambda \right] [n]. 
\ee
The corresponding expression for the emission rate-operator $[R_{ji}]$ is given by
\be
     \Rji [n] =    {4\pi\over hc}
     \int_0^\infty \alpha_{ji}(\lambda)
     \left\{ {2hc^2\over\lambda^5} +\Psi(\lambda) [E(\lambda)][n]\right\} 
     \exp\left(-{hc\over k\lambda T}\right)\,\lambda d\lambda   
\ee
where we have used the definition 
\be
      \Lambda(\lambda)=\Psi(\lambda)/\chi(\lambda),
\ee
and $[E(\lambda)]$ is a {\em linear} operator such that $[E(\lambda)][n]$
gives the emissivity $\eta(\l)$.

Using the rate operator, we can write the rate equations in the form
\be
    \sum_{j<i} n_j \left([R_{ji}][n]+C_{ji}\right)
   &-&n_i\left\{\sum_{j<i} \left(\div{n_j^{*}}{n_i^{*}}\right)
		\left([R_{ij}][n]+C_{ji}\right)\right. \nonumber\\
  &+&\left.\sum_{j>i} 
		 \left([R_{ij}][n]+C_{ij}\right)\right\}      \\
   &\qquad& +\sum_{j>i} n_j 
     \left(\div{n_i^{*}}{n_j^{*}}\right)\left([R_{ji}][n]+C_{ij}\right)
     = 0. \nonumber
\ee
This form shows, explicitly, the non-linearity of the rate equations with
respect to the population densities. Note that in addition, the rate equations
are non-linear with respect to the electron density via the collisional rates.
Furthermore, the charge conservation constraint condition directly couples the
electron densities and the population densities of all level of all atoms and
ions with each other.

In analogy to the operator splitting method discusses above, we split the rate
operator, by writing $\Rij=\Rijstar + (\Rij-\Rijstar)\equiv \Rijstar+\DRij$
(analog for the downward radiative rates), where $\Rijstar$ is the
``approximate rate-operator''. We then rewrite the rate $R_{ij}$ as 
\be
     R_{ij} = \Rijstar [n_{\rm new}] + \DRij [n_{\rm old}]  \label{aro}
\ee
and analogously for the downward radiative rates. In Eq.~\ref{aro}, 
$[n_{\rm old}]$
denotes the current (old) population densities, whereas $[n_{\rm new}]$ 
are the updated (new) population densities to be calculated. The $\Rijstar$ and
$\Rjistar$ are linear functions of the population density operator $[n_k]$ of
any level $k$, due to the linearity of $\eta$ and the usage of the $\Psi$-operator
instead of the $\Lambda$-operator. 

If we insert Eq.~\ref{aro} into Eq.~\ref{rateeqs}, we obtain the following
system for the new population densities:
\be
   & \sum_{j<i} n_{j,\rm new} [R_{ji}^{*}][n_{\rm new}]
   -n_{i,\rm new}\left\{\sum_{j<i} \left(\div{n_j^{*}}{n_i^{*}}\right)
		[R_{ij}^{*}][n_{\rm new}]
	     +\sum_{j>i} 
		 [R_{ij}^{*}][n_{\rm new}]\right\}  \nonumber    \\
   &\quad +\sum_{j>i} n_{j,\rm new} 
     \left(\div{n_i^{*}}{n_j^{*}}\right)[R_{ji}^{*}][n_{\rm new}] 
    +\sum_{j<i} n_{j,\rm new} \left(\DRji[n_{\rm old}]+C_{ji}\right)\label{keyeqn}\\
   &\quad -n_{i,\rm new}\left\{\sum_{j<i} \left(\div{n_j^{*}}{n_i^{*}}\right)
		\left(\DRij[n_{\rm old}]+C_{ji}\right)
    +\sum_{j>i} 
		 \left(\DRij[n_{\rm old}]+C_{ij}\right)\right\}  \nonumber    \\
   &\quad +\sum_{j>i} n_{j,\rm new} 
     \left(\div{n_i^{*}}{n_j^{*}}\right)\left(\DRji[n_{\rm old}]+C_{ij}\right)
     = 0.\nonumber
\ee

Due to its construction, the \Rijstar-operator contains information
about the influence of a particular level on {\it all\/} radiative
transitions. Therefore, we are able to treat the complete multi-level
non-LTE radiative transfer problem including active continua and
overlapping lines.  The $[E(\lambda)]$-operator, at the same time,
gives us information about the strength of the coupling of a radiative
transition to all levels that are considered. This information may be
used to include or neglect certain couplings {\it dynamically} during
the iterative solution of Eq.~\ref{keyeqn}. Furthermore, we have not
yet specified either a method for the formal solution of the radiative
transfer equation or a method for the construction of the approximate
$\L$-operator (and, correspondingly, the $\Rijstar$-operator). We proceed
by considering rapidly expanding spherically symmetric media and use
the tri-diagonal ALO given by Hauschildt \cite{phhs392}. However, any
method for the formal solution of the radiative transfer equation and
the construction of the ALO may be used, including multi-dimensional
and/or time dependent methods.

\subsubsection{Solution of the statistical equations}

The system Eq.~\ref{keyeqn} for $[n_{\rm new}]$ is non-linear with
respect to the $n_{i,\rm new}$ and $n_e$ because the coefficients of the
$\Rijstar$ and $\Rjistar$-operators are quadratic in $n_{i,\rm new}$
and the dependence of the Saha-Boltzmann factors and the collisional
rates on the electron density, respectively.  The system is closed by the
abundance and charge conservation equations.  To simplify the iteration
scheme, and to take advantage of the fact that not all levels strongly
influence all radiative transitions, we use a linearized and splitted
iteration scheme for the solution of Eq.~\ref{keyeqn}.  This scheme has
the further advantage that many different elements in different ionization
stages and even molecules can be treated consistently. A problem where
this is important is the modeling of nova and supernova atmospheres,
where there are typically very large temperature gradients within the
line forming region of the atmosphere.

  To linearize Eq.~\ref{keyeqn}, we follow \cite{rybhum91} and replace
terms of the form 
$n_{j,\rm new} [R_{ji}^{*}][n_{\rm new}]$ in Eq.~\ref{keyeqn} by
$n_{j,\rm old} [R_{ji}^{*}][n_{\rm new}]$: 
\be
    &\sum_{j<i} n_{j,\rm old} [R_{ji}^{*}][n_{\rm new}]
   - n_{i,\rm old}\left\{\sum_{j<i}\left(\div{n_j^{*}}{n_i^{*}}\right)
		[R_{ij}^{*}][n_{\rm new}]
	     +\sum_{j>i} 
		 [R_{ij}^{*}][n_{\rm new}]\right\} \nonumber    \\ 
   &\quad +\sum_{j>i} n_{j,\rm old} 
     \left(\div{n_i^{*}}{n_j^{*}}\right)[R_{ji}^{*}][n_{\rm new}] 
    +\sum_{j<i} n_{j,\rm new} \left(\DRji[n_{\rm
    old}]+C_{ji}\right)\label{subsrate}  \\
   &\quad -n_{i,\rm new}\left\{\sum_{j<i}\left(\div{n_j^{*}}{n_i^{*}}\right)
		\left(\DRij[ n_{\rm old}]+C_{ji}\right)
	     +\sum_{j>i} 
		 \left(\DRij[n_{\rm old}]+C_{ij}\right)\right\} \nonumber \\
   &\quad +\sum_{j>i} n_{j,\rm new} 
     \left(\div{n_i^{*}}{n_j^{*}}\right)\left(\DRji[n_{\rm old}]+C_{ij}\right)
     = 0. \label{linaro}
\ee
This removes the major part of the non-linearity of Eq.~\ref{keyeqn} but the
modified system is still non-linear with respect to $n_e$ and still has the
high dimensionality of the original system. However, as has been noted before,
not all levels are strongly coupled to all other levels.  Equation~\ref{linaro}
can be solved for each element (or groups of elements if they are coupled
tightly) {\it separately\/} if the electron density is
given. Therefore, we split the electron density calculation from the rate
equation solution so that the $n_e$ can be considered as given during the rate
equation solution process and changes in the electron density are then
accounted for in an outer iteration to find a consistent solution of the rate
equations and the electron densities. 

The most important advantage of this method is that it requires the solution of
large {\it linear\/} systems and low-dimensional non-linear system (for the electron
density). Thus, its solution is more stable and uses much less computer
resources (time and memory) than the direct solution of the original non-linear
equations.  This allows us to treat  many more levels with this method then
with more conventional algorithms.  Using a nested iteration scheme like the
one described here will slow down the convergence of the iterations, but this
is more than offset for by the possibility of calculating much larger models
with less memory. Since we are able to solve a separate equation for each group
of elements, we can trivially parallelize the solution by distributing the
groups among the available processors.
Convergence acceleration methods can in principle be used, but they frequently
lead to convergence instabilities in the nested iterations for the solution of
the statistical equilibrium equations.

We have  so far assumed that the electron density $n_e$ is
given. Although this is a good assumption if only trace elements are
considered, the electron density may be sensitive to non-LTE
effects. This can be taken into account by using either a fixed point
iteration scheme for the electron density or, if many species or
molecules are included in the non-LTE equation of state, by a
modification of the LTE partition functions to include the effects of
non-LTE in the ionization equilibrium. The latter method replaces the
partition function, $Q=\sum g_i\exp(-E_i/kT)$, with its non-LTE
generalization, $Q_{\rm NLTE} = \sum b_i g_i\exp(-E_i/kT)$, and uses
$Q_{\rm NLTE}$ in the solution of the ionization/dissociation
equilibrium equation. We use this method because of the
large number of elements with various ionization stages as well
as molecules and condensation of dust grains included in 
statistical equilibrium calculations (and not all of them in non-LTE). 

Our iteration scheme for the solution of the multi-level non-LTE problem
can be summarized as follows: (1) for given $n_i$ and $n_e$, solve the
radiative transfer equation at each wavelength point
and update the radiative rates and the
approximate rate operator, (2) solve the linear system Eq.~\ref{linaro}
for each group for a given electron density, (3) compute new electron
densities (by either fixed point iteration or the generalized partition
function method), (4) if the electron density has not converged to
the prescribed accuracy, go back to step 2, otherwise go to step 1. The
iterations are repeated until a prescribed accuracy for the $n_e$ and the
$n_i$ is reached. It is important to account for coherent 
scattering processes during the solution of the wavelength
dependent radiative transfer equation, it explicitly removes a global
coupling from the iterations.

\subsection{Temperature correction procedure}

In the outermost level of the nested iteration scheme we also iterate for the
temperature structure of the atmosphere using a generalization of the
Uns\"old-Lucy temperature correction scheme to spherical geometry and NLTE
model calculations.  This has proven to work very well even in extreme NLTE
cases such as nova and supernova atmospheres. The temperature correction
procedure also requires virtually no memory and CPU time overheads. The
Uns\"old-Lucy correction scheme (see Mihalas \cite{mihalas70sa} for a
discussion of this and other temperature correction schemes), uses the global
constraint equation of energy conservation to find corrections to the
temperature that will fulfill energy conservation better than the previous
temperatures.  We have found it to be more stable than a Newton-Raphson
linearization scheme and it allows us to separate the temperature corrections
from the statistical equations discussed above.

To derive the Uns\"old-Lucy correction, one uses the fact that
the {\em ratios} of the wavelength averaged absorption and
extinction coefficients 
\begin{eqnarray}
\kappa_P & = & \left(\int_0^\infty \kappa_\l B_l\,d\l\right)/B \\
\kappa_J & = & \left(\int_0^\infty \kappa_\l J_l\,d\l\right)/J \\
\chi_J & = & \left(\int_0^\infty \chi_\l F_l\,d\l\right)/F \\
\end{eqnarray}
(where $B,J,F$ denote the wavelength integrated Planck function, mean intensity
and radiation flux, respectively) depend much less on values of the
independent variables than do the averages themselves. 

Dropping terms of order $(v/c)$, one can then
use the angular moments of the SSRTE to show that in order to
obtain radiation equilibrium $B$ should be corrected by an amount
\be
\delta B(\tau) &=& \frac{1}{\kappa_P}\big\{\kappa_J J - \kappa_P B+ \dot S/(4
\pi)\big\} \\ &&- \big\{ 2(H(\tau)-H_0(\tau)) - \int_0^\tau q\, \chi_F
\,(H(\tau)-H_0(\tau))\big\},\label{uleqn}\nonumber
\ee
where $H\equiv F/4\pi$, $H_0(\tau)$ is the value of the target luminosity at
that particular depth point (variable due to the velocity terms in the SSRTE
and non-mechanical energy sources, the total {\em observed} luminosity $H_0(0)$
is an input parameter), Here, $q$ is the ``sphericity factor'' given by \[ q =
\frac{3f - 1}{rf}, \] where $f(\t)=K(\t)/J(\t)$ is the ``Eddington
factor'' and $K=\int 
\mu^2 I \, d\mu$ is the second angular moment of the mean intensity. $\dot S$
describes all additional sources of energy such as mechanical energy supplied
by winds or non-thermal ionization due to $\gamma$--ray deposition. 

The first term in Eq.~\ref{uleqn} corresponds simply to a $\L$ iteration term
and will thus provide too small temperature corrections in the {\em inner}
parts of the atmosphere (but work fine in the outer, optically thin parts). The
second term of Eq.~\ref{uleqn}, however, is the dominant term in the inner
parts of the atmosphere. It provides a very good approximation to the
temperature corrections $\Delta T$ deep inside the atmosphere. Following
\cite{unsoeld55}, we found that it is sometimes better to modify this general
scheme by, e.g., excluding the contributions of extremely strong lines to the
opacity averages used in the $\Delta T$ calculations because they tend to
dominate the average opacity but do not contribute as much to the total error
in the energy conservation constraint. 

\subsection{Global iteration scheme}

As the first step in our outermost iteration loop (the ``model iteration'')
we use the current best guess of $\{T,n_i\}$ as function of radius
to solve the hydrostatic or hydrodynamic equations to calculate an 
improved run of $\Pgas$ with radius. Simultaneously, the population numbers
are updated to account for changes in $\Pgas$. The next major step is
the computation of the radiation field for each wavelength point (the 
``wavelength loop''),
which has the prerequisite of a spectral line selection procedure for
LTE background lines. Immediately after the radiation field at any given
wavelength is known, the radiative rates and the rate operators are updated
so that their calculation is finished after the last wavelength point. 
In the next steps, the population numbers are updated by solving the 
rate equations for each NLTE species and new electron densities 
are computed, this gives improved estimates for $\{n_i\}$. The last part of the 
model iteration is the temperature correction scheme outlined above (using
opacity averages etc.\ that were computed in the wavelength loop) which 
delivers an improved temperature structure. If the errors in the constraint
equations are larger than a prescribed accuracy, the improved
$\{T,n_i\}$ are used in another model iteration. Using this scheme,
about 10--20 model iterations are typically required to reach convergence
to better than about 1\% relative errors, depending on the quality of
the initial guess of the independent variables and the complexity of
the model.

\section{Parallelization}

Solving the above set of coupled non-linear equations for large numbers
of NLTE species requires large amounts of memory to store the rates
for each level in all the model atoms at each radial grid point, and
large amounts of CPU time because many wavelength points are required
in order to resolve the line profiles in the co-moving frame.  In order
to minimize both CPU and memory requirements we have parallelized the
separate {\tt Fortran 90} modules which make up the {\tt PHOENIX} code.
Our experience indicates that only the simultaneous use of data and task parallelism
can deliver reasonable parallel speedups~\cite{hbapara97}.
This involves:
\begin{enumerate}
\item The radiative transfer calculation itself, where we divide up
the characteristic rays among nodes and use a ``reduce'' operation to
collect and send the $\jnu$ to all the radiative transfer and NLTE rate
computation tasks (data parallelism); 
\item the line opacity which requires
the calculation of up to 50,000 Voigt profiles per wavelength point at
each radial grid point, here we split the work amongst the processors
both by radial grid points and by dividing up the individual lines to be
calculated among the processors (combined data and task parallelism); and
\item the NLTE calculations.  The NLTE calculations involve three separate
parts: the calculation of the NLTE opacities, the calculation of the rates at
each wavelength point, and  the solution of the NLTE rate and statistical
equilibrium equations.  To prevent communication overhead, each task computing
the NLTE rates is forced to be on the same node with the corresponding task
computing NLTE opacities and emissivities, (combined data and task
parallelism). The solution of the rate equations parallelizes trivially with
the use of a diagonal approximate rate operator.
\end{enumerate}

In the latest version of our code, {\tt PHOENIX 9.1}, we have incorporated
the additional strategy of distributing each NLTE species (the total
number of ionization stages of a particular element treated in NLTE) on
separate nodes. Since different species have different numbers of levels
treated in NLTE (e.g. Fe~II [singly ionized iron] has 617 NLTE levels,
whereas H~I has 30 levels), care is taken to balance the number of
levels and NLTE transitions treated on each  node to avoid unnecessary
synchronization and communication problems. We have also parallelized the selection of
background atomic and molecular LTE lines (a significant amount of work
considering that our combined line lists currently include about 400
million lines and we expect line lists with about 1 billion lines in
the near future).  Although the line selection seems at first glance to
be an inherently serial process, since a file sorted in wavelength with
selected lines must be written to disk, we are able to obtain reasonable
speedups, by  employing a client-server model with a server line-selection
task which receives the selected lines and writes them to disk and client
nodes which read pieces (blocks) of the line list files and carry out the 
actual selection processes on each block of lines.

In addition to the combined data and task parallelism discussed above,
\phoenix\ also uses simultaneous explicit task parallelism by allocating
different tasks (e.g., atomic line opacity, molecular line opacity,
radiative transfer) to different nodes. This can result in further
speed-up and better scalability but requires a careful analysis of
the workload between different tasks (the workload is also a function
of wavelength, e.g., different number of lines that overlap at each
wavelength point) to obtain optimal load balancing.

\subsection{Wavelength Parallelization}

The parallelization of the computational workload outlined in the previous
paragraphs requires synchronization between the radiative transfer tasks and
the NLTE tasks, since the radiation field and the $\lstar$
operator must be passed between them. In addition, our standard
model calculations use 50 radial grid points and as the number of nodes
increases, so too does the communication and loop overhead, therefore,
pure data parallelism does not deliver good scalability. We found good
speedup up to about 5 nodes for a typical calculation, with
the speedup close to the theoretical maximum.  However, for 5 nodes the
communication and loop overheads begin to become significant and it is
not economical to use more than 10 nodes (depending on the machine and
the model calculation, it might be {\em necessary} to use more nodes to
fit the data in the memory available on a single node).

Since the number of wavelength points in a calculation is very large
and the CPU time scales linearly with the number of wavelength points, 
parallelization with respect to the wavelength points can lead
to large speedups and to the ability to use very large numbers of nodes
available on massively parallel supercomputers. This poses no difficulties
for static configuration, but the coupling of the wavelengths points
in expanding atmospheres makes the wavelength parallelization much more
complex.

We have developed a wavelength parallelization based on a toroidal 
topology that uses 
the concept of wavelength ``clusters'' to distribute a set of wavelength
points (for the solution of the wavelength dependent radiative transfer)
onto a different set of nodes, see Fig.~\ref{design} \cite{bhpar298}.
In order to achieve optimal load balance and, more importantly, in order
to minimize the memory requirements, each cluster (a column of
nodes indicated in Fig.~\ref{design}) works on a single
wavelength point at any given time. Each cluster can consist of a {\em number}
of ``worker'' nodes where the worker node group uses parallelization methods
discussed above (see also Ref.~\cite{hbapara97}). In order to avoid
communication 
overhead, we use {\em symmetric}  wavelength clusters:
each ``row'' of worker nodes in Fig.~\ref{design} performs identical
tasks but on a different set of wavelength points for each cluster. We thus
arrange the total number of nodes $N$ in a rectangular matrix with $n$
columns and $m$ rows, where $n$ is the number of clusters and $m$ is
the number of workers for each cluster, such that $N=n * m$. Another
way of visualizing this parallelization technique is to consider each
wavelength cluster as a single entity (although not a single node or CPU)
that performs a variety of different tasks at each wavelength
point. The entity (cluster) itself is then further subdivided into individual
nodes or CPUs each of which perform a given set of tasks {\em at a
particular wavelength point}. This topology can be implemented 
very efficiently in the context of the {\tt MPI} library, see \cite{hbapara97}
for details.

For a static model atmosphere, all wavelengths and thus wavelength
clusters are completely independent and execute in parallel with
{\em no} immediate communication or synchronization along the rows
of Fig.~\ref{design}.  Communication is only necessary {\em after}
the calculation is complete for all wavelengths points on all nodes to
collect, e.g., the rates and rate operators. Therefore, the speedup is
close (80\%) to the theoretical maximum, limited only by to combined IO
bandwidth of the machine used.

In order to parallelize the spectrum calculations for a model atmosphere with a
{\em global velocity field}, such as the expanding atmospheres of novae,
supernovae or stellar winds, we need to take the mathematical character of the
RTE into account. For monotonic velocity fields, the RTE is an initial value
problem in wavelength (with the initial condition at the smallest wavelength
for expanding atmospheres and at the largest wavelength for contracting
atmospheres). This initial value problem must be discretized fully implicitly
to ensure stability.  In the simplest case of a first order discretization, the
solution of the RTE for wavelength point $i$ depends only on the results of the
point $i-1$. In order to parallelize the spectrum calculations, the wavelength
cluster $n_i$ computing the solution for wavelength point $i$ must know the
specific intensities from the cluster $n_{i-1}$ computing the solution for
point $i-1$. This suggests a ``pipeline'' solution to the wavelength
parallelization, transforming the ``matrix'' arrangement of nodes into a
``torus'' arrangement where data are sent along the torus' circumference. Note
that only the solution of the RTE is affected by this, the calculation of the
opacities and rates remains independent between different wavelength clusters.
In this case, the wavelength parallelization works as follows: Each cluster can
independently compute the opacities and start the RT calculations (e.g., the
$\lstar$ calculations, hereafter called the {\em pre-processing phase}), it
then waits until it receives the specific intensities for the previous
wavelength point, then it finishes the solution of the RTE and {\em
immediately} sends the results to the wavelength cluster calculating the next
wavelength point (to minimize waiting time, this is done with non-blocking
send/receives), then proceeds to calculate the rates etc.\ (hereafter called
the {\em post-processing phase} and the new opacities for its {\em next}
wavelength point and so on.

The important point in this scheme is that each wavelength cluster can
execute the post-processing phase of its current wavelength point and
pre-processing phase of its next wavelength point {\em independently
and in parallel with all other clusters}. This means that the majority
of the total computational work can be done in parallel, leading to a
substantial reduction in wall-clock time per model.  Ideal load balancing
can be obtained by dynamically allocating wavelength points to wavelength
clusters. This requires only primitive logic with no measurable overhead,
however it requires also communication and an arbitration/synchronization
process to avoid deadlocks.  Typically, the number of clusters $n$
(4-64) is much smaller than the number of wavelength points, $n_{\rm
wl}\approx 300,000$, so that at any given time the work required for each
wavelength point is roughly the same for each cluster (the work changes
as the number of overlapping lines changes, for example). Therefore,
a simple {\em round robin} allocation of wavelength points to clusters
(cluster $i$ calculates wavelength points $i$, $n+i$, $2n+i$ and so
on) can be used which will result in nearly optimal performance if the
condition $n \ll n_{\rm wl}$ is fulfilled. However, once the pipeline
is full, adding further wavelength clusters cannot decrease the 
time required for the calculations, setting a limit for the efficient
``radius'' of the torus topology. However, this limit can be increased
somewhat by increasing the number of worker nodes per wavelength cluster.

\subsubsection{Scaling Results}

For a simple supernova test calculation, we examine both the scaling and
performance tradeoff of spatial versus wavelength parallelization.
Figure~\ref{sn} presents the results of our timing tests for one iteration of a
Type Ic supernova model atmosphere, with a model temperature $\Tmod = 12,000$~K
(the observed luminosity is given by $L=4\pi R^2 \Tmod^4$), characteristic
velocity $\vno=10000$~\kms, 4666 NLTE levels, 163812 NLTE lines, 211680 LTE
lines (for simplicity, all line profile were assumed to be Gaussian),
non-homogeneous abundances, and 260630 wavelength points. This is a typical
test for production calculations and we have designed this test to have the
highest potential for synchronization, I/O waiting, and swapping to reduce
performance to simulate a worst case scenario for the parallel performance. It
is however, characteristic of the level of detail needed to accurately model
supernovae. This calculation has also been designed to barely fit into the
memory of a single node. The behavior of the speedup is very similar to the
results obtained for test case using a model of a nova explosion
\cite{bhpar298}.  The ``saturation point'' at which the wavelength pipeline
fills and no further speedup can be obtained if more wavelength clusters are
used for the machines used here, occurs at about 5 to 8 clusters. More clusters
will not lead to larger speedups, as expected. Larger speedups can be obtained
by using more worker nodes per cluster, which also drastically reduces the
amount of memory required on each node.

\section{Discussion and Conclusions}

We have presented  our approach to the  numerical solution to the generalized
stellar atmosphere problem in the presence of rapidly expanding flows. We have
shown how the use of accelerated $\Lambda$ operators may result in the
formulation of the problem in such a way that extremely detailed model atoms
may be handled in NLTE and the problem can be parallelized in a way that
significantly reduces the per processor memory and CPU requirements with modest
communication overhead. Parallelization also allows much more complex models to
be calculations by giving us access to the large memory sizes that are
available on modern parallel supercomputers. Currently, our largest model
calculations involve 6000 atomic NLTE level with 65000 primary NLTE lines that
are modeled individually, 2-10 million weak atomic secondary NLTE and LTE
background lines and, for models of cool stellar winds, 150 million molecular
lines. Simulations of this size and level of detail were simply not possible
before the development of new radiative transfer algorithms and the
availability of parallel supercomputers.  We believe that the next step --- the
computation of moving flows in three spatial dimensions, is becoming tractable
on modern parallel supercomputers. There continues to be an urgent need for
improvements in the fundamental atomic data which serves as input to these
calculations.

\begin{ack}
We thank our many collaborators who have contributed to the development of {\tt
PHOENIX}, in particular we would like to thank France Allard, David Branch,
Steve Shore, Sumner Starrfield, Jason Aufdenberg, and Andreas Schweitzer. This
work was supported in part by NASA ATP grant NAG 5-3018 and LTSA grant NAG
5-3619 and NSF grant AST-9720804 to the University of Georgia, and by NSF grant
AST-9417242, NASA grant NAG5-3505 and an IBM SUR grant to the University of
Oklahoma.  Some of the calculations presented in this paper were performed on
the IBM SP2 and SGI Origin 2000 of the UGA UCNS, at the San Diego Supercomputer
Center (SDSC), the Cornell Theory Center (CTC), and at the National Center for
Supercomputing Applications (NCSA), with support from the National Science
Foundation, and at the NERSC with support from the DoE. We thank all these
institutions for a generous allocation of computer time.
\end{ack}

\bibliography{refs,rte,crossrefs}

\clearpage

\begin{figure}
\vskip -2cm
\caption{\label{flow1}Relation between (some of) the physical and mathematical
blocks that describe the physics of a stellar atmosphere. In order
to calculate a model atmosphere, a set of value of the physical 
variables, e.g., temperatures, densities, population densities and
the radiation field, must be found that satisfies all constraints
simultaneously.}
\end{figure}

\begin{figure}
\caption{\label{flow2}Relations between the main types of variables
represented by blocks are indicated. The labels name the equations
that relate the block to each other.
}
\end{figure}

\begin{figure}
\caption{The basic ``torus'' design of the wavelength-parallelized version of
\phoenix: groups of processors are divided up into wavelength clusters which
will work on individual wavelength points, each wavelength cluster is further
divided into a set of worker nodes, where each worker node is assigned a set of
specific tasks, e.g., it will work on the LTE background line opacity for a set
of radial points. Our design requires that each worker node on all wavelength
clusters work on exactly the same set of tasks, although additional inherently
serial operations can be assigned to one particular master worker, or master
wavelength cluster. This reduces communication between clusters to its absolute
minimum and allows the maximum speedup.} \label{design} 
\end{figure}

\begin{figure}
\caption{Scalability of the Supernova model atmosphere test run
as function of the number of nodes (processing elements or nodes) used.
The y-axis gives the speedup obtained relative to the serial run. The
different symbols show the results for different numbers of worker tasks
for each wavelength cluster.}\label{sn} 
\end{figure}
\end{document}